\documentstyle[12pt]{article}

\def\be{\begin{equation}}
\def\ee{\end{equation}}
\def\bea{\begin{eqnarray}}
\def\eea{\end{eqnarray}}

\def\det{{\rm det}}
\def\haf{\frac{1}{2}}

\def\tr{{\rm Tr}}
\def\e{{\rm e}}
\def\cf{{\cal F}}
\def\bmn{B_{\mu\nu}}
\def\sq{\sqrt}
\def\dpn{2\pi N}
\def\ap{\alpha'}

\begin{document}

\begin{flushright}
IPM/P-99/058\\
hep-th/9910062
\end{flushright}

\pagestyle{plain}

\begin{center}
{\Large {\bf (m,n)-Strings In IIB Matrix Model}}

\vspace{.5cm}

Shahrokh Parvizi$_{a,c}$ and  Amir H. Fatollahi$_{b,c}$

\vspace{.2 cm}
{\it a) Shahid Rajaee University, Faculty of Science, Physics
Department,}\\
{\it P.O.Box 16785-163, Lavizan, Tehran, Iran}\\
\vspace{.2 cm}

{\it b) Institute for Advanced Studies in Basic Sciences (IASBS),}\\
{\it P.O.Box 159, GavaZang, Zanjan 45195, Iran}\\

\vspace{.2 cm}
{\it c)\footnote{Postal address} Institute for Studies in
Theoretical Physics and Mathematics (IPM),}\\
{\it P.O.Box 19395-5531, Tehran, Iran}
\vspace{.2 cm}

{\sl E-mails: parvizi, fath@theory.ipm.ac.ir}

\vspace{.1 cm}
\begin{abstract}
By adding gauge fields to the D-string classical solution, which have
non-zero contribution to commutators in
continuum limit (extreme large $N$), we introduced
$(m,n)$-strings in IIB matrix model. It is found that the size of matrices
depends on the value of the electric field. The tension of these
strings appears in $SL(2,Z)$ invariant form. The interaction for parallel
and angled strings are found in agreement with the string theory for small
electric fields.
\end{abstract}
\end{center}

Matrix models have been conjectured to be the non-perturbative formulations
of string theories \cite{BFSS,IKKT,DVV}. Among the years these models have
shown their abilities in covering many aspects of string theory \cite{ToBa,AI},
including representation of various BPS configurations, D-brane interactions
and dynamics, compactification subjects and  string dualities. However
there are some issues which have lacked the complete
correspondence between string theory and matrix models. One may recall some
of these issues as the absence of transverse NS5-branes in m(atrix) model
and
the important one, formulation of matrix models in general background
fields.
In an interesting paper \cite{CDS} for compactification of matrix models,
a non-commutative structure was resolved to push formulation of matrix
models in non-zero background form fields. More specifically,
it was shown
that the amount of the non-commutativity in gauge theory of
D-branes is measured by the form field flux.

In string theory side this issue was approached in the context of
bound-states of Fundamental strings (F-strings) with D-branes and
also D-branes with each other. In a series of works
\cite{9709054,ardalans,AAS,9712199,anazawa}
it was understood that this new structure can be derived by a
string theoretical description of D-branes \cite{9510017}; considering
non-trivial form field background on the world-volume of D-branes in
extracting the boundary conditions of open strings ended on D-branes.
It is observed that the non-zero
background causes mixing the Dirichlet and Neumann boundary conditions
at the ends of strings and it is the reason for non-commutative
coordinates on the D-brane world-volume.
As is known in the D-brane action, called DBI action, the combination
$\cf_{\mu\nu}=\bmn-\partial_{[\mu}A_{\nu]}$ appears, with $\bmn$ as
the NS B-field and $A_\mu$ as the gauge field living on the D-brane.
In this picture, non-zero background have the interpretation as the
bound-states of F-strings with D-branes (in the case
$\cf_{0i} \neq 0$)  or D-branes with each other ($\cf_{ij}\neq 0$)
\cite{tasi,calkle}.
So one may interpret the F-string bound-states as turned on electric
field, and D-brane bound-states as turned on
magnetic ones in the gauge theory of D-branes.
In \cite{AAS} the spectrums of these bound-states have been studied to
compare them with duality predictions. In the most famous case FD-string
bound-states, so-called $(m,n)$-strings \cite{schwarz,9510135}, the
tension appeared in the $SL(2,Z)$ invariant form
\bea\label{0}
T_{(m,n)}=\frac{1}{2\pi\alpha'}\sqrt{m^2+\frac{n^2}{g_s^2}}.
\eea

In  \cite{mli,9908141} this issue was considered in the context
of IIB matrix model \cite{IKKT}.
The crucial observation was about those sectors of commutation relations
between the position matrices of D-branes
which remain non-zero in the continuum limit (extreme large-$N$ limit).
It was shown that the non-zero commutation relations between
the position matrices are responsible for the non-commutativity
in the world-volume and gauge theory of D-branes.

It is the purpose of this work to consider the $(m,n)$-strings in the
context of IIB matrix model, as they are expected from the $SL(2,Z)$
symmetry of IIB string theory \footnote{This subject has been addressed in
\cite{sat1}. Also the interaction between a F-string and D-string is
calculated in \cite{sat2} in a compact space.}. Based on the background
mentioned above, we introduce $(m,n)$-string by adding gauge fields to
D-string classical solution, which have non-zero contribution to
commutators in continuum limit (extreme large $N$). It is found that the
size of matrices depends on the value of the electric field. The tension
of these strings appears in $SL(2,Z)$ invariant form. The interaction for
parallel and angled strings are found in agreement with string theory
for small electric fields.

The IIB matrix model action is \footnote{Here we differ with the
original action by an overall numerical factor, to set the tension of a
D-string $\frac{1}{2\pi\alpha' g_s}$.}
\bea\label{1}
S=\frac{1}{\sqrt{6}\pi\alpha'^2g_s}\left(-\frac{1}{4} \tr[X_\mu,X_\nu]^2-
\frac{1}{2}\tr\bar\psi\Gamma_\mu[\psi,X_\mu]\right)+
\frac{\sqrt{3}\pi}{\sqrt{2}g_s}\tr({\bf 1}),
\eea
where $X_\mu$ and $\psi$ are hermitian $N\times N$ matrices
$SO(10)$ bosons and fermions, respectively.
The classical equations of motion read as
\bea\label{5}
\sum_\mu [X_\mu,[X_\mu,X_\nu]]=0.
\eea
Solutions with $[X_\mu,X_\nu]\sim 1$ and
with the other $X$'s vanishing are in special interest to represent D$p$-branes
of string theory. For D-string we have
\bea\label{10}
X_0 = \frac{qT}{\sq{2\pi N}},\;\;\; X_1=\frac{pL}{\sq{\dpn}},\;\;\;
X_{i\geq 2}=0,
\eea
where $q$ and $p$ are matrices with size $N$ to be large, with
the commutation relation and uniformly eigenvalue distributions as:
\bea\label{15}
[q,p]=i,\;\;\;\;\; 0\leq q,\;p \leq \sq{\dpn}.
\eea

The one-loop effective action $W$ is calculated in \cite{IKKT} with
the bosonic, fermionic and ghost contributions:
\bea\label{20}
W=-\ln\bigg(
\det^{-\haf}(P_\lambda^2\delta_{\mu\nu}-2iF_{\mu\nu})\cdot
\det^{\haf}(\partial_t+\sum_{i=1}^{9}P_i\gamma_i)\cdot
\det(P_\lambda^2)\bigg),
\eea
with $P_\mu*=[X_\mu,*]$, $F_{\mu\nu}*=[f_{\mu\nu},*]$,
 and $f_{\mu\nu}=i[X_\mu,X_\nu]$.
In the cases of our interest we have $[X_\lambda, f_{\mu\nu}]=c-number$, and
so $P_\lambda^2$ and $F_{\mu\nu}$ are simultaneously diagonalizable.
Writing $F_{\mu\nu}$ in the Jordan form introduced in \cite{IKKT} with
$a_1, ..., a_5$ as the off-diagonal entries one finds for the one-loop
effective action (\ref{20}) the form
\bea\label{105}
W&=&\frac{1}{2}\sum_{i=1}^5 \tr\log \left(1-\frac{4a_i^2}{(P_\lambda^2)^2}
\right)\nonumber\\
&~&-\frac{1}{4}\sum_{s_1,...,s_5=\pm 1; s_1...s_5=1}
\tr\log\left(1-\frac{a_1s_1+...+a_5s_5}{P_\lambda^2}\right).
\eea
It is worth mentioning that the configurations with $F_{\mu\nu}\equiv 0$
for all $\mu,\nu$, like the same as D-string (\ref{10}),
have vanishing quantum corrections due to the algebra
\bea\label{25}
W\sim \tr\log(P_\lambda^2)\;(\frac{10}{2}- \frac{16}{4}- 1)=0.
\eea
This is an evidence for the stability of the solution under quantum fluctuations, as one
expects for a BPS state.

Also, let us calculate the tension
of our D-string solution. By inserting the solution (\ref{10}) in the action one
finds:
\bea\label{30}
S=\frac{1}{2\sq{6}\pi\ap^2g_s}\frac{T^2L^2}{4\pi^2}\frac{1}{N}+
\frac{\sq{3}\pi}{\sq{2}g_s}N.
\eea
The equation of motion for the size of matrices, $N$, gives \cite{IKKT}
\bea\label{35}
\frac{\partial S}{\partial N}=0 \Rightarrow N=\frac{1}{\sq{6}\pi\ap}\frac{TL}{2\pi},
\eea
which by re-inserting it in the on-shell action (\ref{30}) one has
\bea\label{40}
S=\frac{1}{2\pi\ap g_s}\cdot TL,
\eea
\noindent The tension of D-string reads therefore $\frac{1}{2\pi\ap g_s}$.
For the solutions
with the interpretation of $n$ number of D-strings one can simply 
multiply
the coordinates by ${\bf 1}_n$ as
\bea\label{45}
X_\mu\rightarrow X_\mu\otimes {\bf 1}_n.
\eea
The on-shell action is then given by
\bea\label{50}
S=\frac{n}{2\pi\ap g_s}\cdot TL,
\eea
with the clear interpretation for the tension of the
bound-state of $n$ D-strings.

It is known that the bound-state of D-strings and F-strings,
called $(m,n)$-strings, are described by turning on the electric
field in the D-string gauge theory world-volume \cite{tasi,calkle}.
So one should add non-trivial bundles to the solutions \cite{9803166}:
\bea\label{55}
X_0 = \frac{qT}{\sq{2\pi N}}+D_0,\;\; X_1=\frac{pL}{\sq{\dpn}}+D_1,
\;\;X_{i\geq 2}=0,
\eea
setting the commutation relation to be
\bea\label{60}
f_{01}=i[X_0,X_1]=-\frac{TL}{\dpn}-c,
\eea
with $c$ independent of $N$. By inserting this solution in the action one finds:
\bea\label{65}
S=\frac{1}{2\sq{6}\pi\ap^2g_s}\frac{T^2L^2}{4\pi^2}\frac{1}{N}+
c^2N+\frac{TL}{\pi}c+\frac{\sq{3}\pi}{\sq{2}g_s}N.
\eea
Notice that there is a term independent of $N$; it is the same topological term known to
be added to the DBI action for D-branes.
The equation of motion for the size of matrices gives
\bea\label{70}
N=\frac{1}{\sq{6\pi^2\ap^2+c^2}}\frac{TL}{2\pi}.
\eea
By re-inserting this eq. in the on-shell action (\ref{65}) one obtains
\bea\label{75}
S=\frac{1}{2\pi\ap g_s}\frac{\sq{6\pi^2\ap^2+c^2}}{\sq{6}\pi\ap}
\cdot TL+\frac{c}{2\sq{6}\pi^2\ap^2g_s}\cdot TL.
\eea
To set the tension of this object as the same of $(m,1)$-string of
formula (\ref{0}) one adjusts\footnote{The quantization of electric field
$\cf_{0i}$ is understood as the momentum of gauge field $A_i$ in
the compact direction \cite{9510135,calkle}.}
\bea\label{80}
c=\sq{6}\pi\ap\cdot \cf,\;\;\;\;\;\cf\equiv mg_s,
\eea
and finds the on-shell action
\bea\label{85}
S=\frac{TL}{2\pi\ap }\sq{\frac{1}{g_s^2}+m^2}+\frac{TL}{2\pi\ap}\cdot m.
\eea
The tension appeared in the first term is the same as $(m,1)$-string given
in
the expression (\ref{0}). The $g_s$ independent term is here sitting as
the $N$ independent term.
This term is known as one which is responsible for the
coupling of RR field to D-brane (\cite{tasi}, p33-35).
Finally one should specify the bundles $D_\mu$. In a specific
gauge one chooses \cite{9803166}
\bea\label{90}
D_0\equiv 0,\;\;\;D_1\equiv \frac{p\sq{\dpn}}{T}\cdot c,\;\;\;D_{i\geq2}
\equiv 0.
\eea
Construction of $(m,n)$-strings can be obtained as the same as transformation
(\ref{45}), to find the tension (\ref{0}).

In the case of pure D-strings $(m=0)$ it is known how to introduce two or
more separated of them \cite{IKKT}, but this is more difficult in
the case
of $(m,n)$-strings. As we see from (\ref{70}) the number of F-strings, $m$,
enters in the size of matrices. So one should choose the sizes of
matrices different. Two separated $(m_1,1)$- and $(m_2,2)$-strings
may be presented by
\bea\label{91}
X_0&=&\left(
\begin{array}{cc}
\frac{q_1T}{\sq{2\pi N_1}}+D_{01} & 0 \\
 0 & \frac{q_2T}{\sq{2\pi N_2}}+D_{02} \\
\end{array}
\right), \;\;
X_1 = \left(
 \begin{array}{cc}
\frac{p_1L}{\sq{2\pi N_1}}+D_{11} & 0 \\
 0 & \frac{p_2L}{\sq{2\pi N_2}}+D_{12} \\
 \end{array}
\right), \nonumber\\
X_2&=& \frac{r}{2}\sigma_3,\;\;\;X_{i\geq 3}=0,
\eea
with distance $r$ in 2nd direction, accompanied by the commutation
relations as
\bea\label{92}
[q_a,p_b]=i \delta_{ab},\;a,b=1,2,\;
f_{01}=\left(
\begin{array}{cc}
\frac{-TL}{2\pi N_1}-c_1&0\\
0& \frac{-TL}{2\pi N_2}-c_2\\
\end{array}
\right),
\eea
and $\sigma_3$ as the 3rd Pauli matrix.
Again by inserting the solution in the action and solving the equation of
motion for $N_{1,2}$ one finds
\bea\label{93}
N_{1,2} =\frac{1}{\sq{6\pi^2\ap^2+c_{1,2}^2}}\frac{TL}{2\pi}.
\eea
By re-inserting this eq. in the action and choosing
\bea\label{94}
c_{1,2} =\sq{6}\pi\ap\cdot \cf_{1,2}, \;\;\;\cf_{1,2} \equiv m_{1,2}g_s,
\eea
one finds the desired on-shell action
\bea
S=\frac{TL}{2\pi\ap }\sq{\frac{1}{g_s^2}+m_1^2}+\frac{TL}{2\pi\ap}\cdot m_1
+\frac{TL}{2\pi\ap }\sq{\frac{1}{g_s^2}+m_2^2}+\frac{TL}{2\pi\ap}\cdot m_2.
\eea
As is seen from (\ref{93}) the sizes of matrices are corrected in the
small $\cf$ limit by $\sim O(\cf^2)$.

The non-trivial check of the above picture for $(m,n)$-strings
 is interaction considerations.

{\it Interaction Of Two Parallel $(m_1,1)$- And $(m_2,1)$-Strings}

Since in the parallel case only the difference of $m_1$ and $m_2$ enters
in the force, we put $m=m_1=-m_2$ for simplicity. We obtain 
the following solution for two strings in distance $r$ in 2nd direction, 
\bea\label{95}
X_0 &=& \frac{qT}{\sq{2\pi N}}{\bf 1}_2,\;\;
X_1=\frac{pL}{\sq{\dpn}}{\bf 1}_2+\frac{p\sq{\dpn}}{T}c\sigma_3,\nonumber\\
X_2&=&\frac{1}{2}r\sigma_3,\;\;X_{i\geq 3}=0.
\eea
Here $c=\sq{6}\pi\ap\cdot \cf$. So one finds
\bea\label{100}
f_{01}&=&\frac{-TL}{\dpn}{\bf 1}_2-c\sigma_3 \Rightarrow
F_{01}=i[P_0,P_1]=-c\Sigma_3,\nonumber\\
&~&{\rm otherwise}\;\;\;f_{\mu\nu}=0,\;\;\;P_2=\frac{r}{2}\Sigma_3,
\eea
where $\Sigma_3 *=[1\otimes\sigma_3,*]$. $\Sigma_3$ has 2, -2, 0 and 0 as
eigenvalues. So one finds the eigenvalues of $P_\lambda^2$ as
$2(2c)(k+\frac{1}{2})+r^2$, with $k$ for the oscillator number. The traces
can be performed and according to the notation of \cite{9703038} one
finds
\bea
W\sim \int_0^\infty \frac{ds}{s} \e^{-sr^2}
\frac{\cosh(4cs)-4\cosh(2cs)+3}{\sinh(2cs)}.
\eea
By changing the integral variable $s'=cs$, for large
separation strictly we have
\bea\label{110}
W\sim -8Nc^3\frac{1}{r^6}+\cdots = -8 \frac{6\pi^2 \ap^2 TL}{2\pi}
\frac{\cf^3}{\sq{1+\cf^2}}\frac{1}{r^6}+\cdots .
\eea
This result coincides with the string theory result (\cite{9712199} Eq. 17)
for small $\cf$.
This restricted correspondence between matrix model and string theory
calculations in this case has been already seen in a T-dual version.
As we are in IIB string theory with $(m,n)$-strings, the T-dual picture
exists in IIA with moving D0-branes with the speed $v=mg_s=\cf$
\cite{9709054,9712199}.
So the interaction we calculated between $(m,1)$ and $(-m,1)$
strings corresponds to interaction between two D0-branes moving with
speeds $mg_s$ and $-mg_s$ toward each others. In
m(atrix) model side the phase shift $\delta$ of scattering corresponds to
$W$ \cite{9705091} to be
\bea
\delta\sim v^3 \frac{1}{r^6},
\eea
from which the famous result $v^4/r^7$ is found for the potential via
eikonal
approximation. Now, this result of m(atrix) model coincides with
string theory only for small $v$ \cite{smallv}.

{\it Interaction Of Two Angled $(m_1,1)$- And $(m_2,1)$-Strings}

Here as the parallel case we set $m=m_1=-m_2$ \footnote{It can be shown that
in the small $\cf$ limit, again like the parallel case, the difference of
$m_1$ and $m_2$ appears in the interaction.}.
The configuration with two angled string is presented by
\bea\label{115}
X_0&=& \frac{qT}{\sq{2\pi N}}{\bf 1}_2,\;\;
X_1=\left(\frac{pL}{\sq{\dpn}}{\bf 1}_2+
\frac{p\sq{\dpn}}{T}c\sigma_3\right)
\left(
 \begin{array}{cc}
 \cos \frac{\theta}{2} & 0 \\
 0 &  \cos \frac{\theta}{2} \\
 \end{array}
 \right),  \nonumber\\
X_{2}&=&\left(\frac{pL}{\sq{\dpn}}{\bf 1}_2+
\frac{p\sq{\dpn}}{T}c\sigma_3\right)
\left(
 \begin{array}{cc}
 \sin \frac{\theta}{2} & 0 \\
 0 &  -\sin \frac{\theta}{2} \\
 \end{array}
 \right), \nonumber\\
X_3&=&\frac{r}{2}\sigma_3,\;\;   X_{i\geq 4}=0.
\end{eqnarray}
So one finds
\bea\label{120}
f_{01}&=&\frac{-TL}{\dpn}{\bf 1}_2 \cos\frac{\theta}{2}
-c \sigma_3 \cos \frac{\theta}{2} \Rightarrow
F_{01}=i[P_0,P_1]=-c \Sigma_3 \cos \frac{\theta}{2},\nonumber\\
f_{02}&=&\frac{-TL}{\dpn}{\bf 1}_2 \sin\frac{\theta}{2}
-c\sigma_3 \sin\frac{\theta}{2} \Rightarrow
F_{02}=i[P_0,P_2]=-\frac{TL}{\dpn}\Sigma_3 \sin\frac{\theta}{2},\nonumber\\
&~&{\rm otherwise}\;\;\;f_{\mu\nu}=0,\;\;\;P_3=\frac{r}{2}\Sigma_3.
\eea
We define $\alpha\equiv 2c \cos \frac{\theta}{2}$ and
$\beta\equiv 2\frac{TL}{\dpn} \sin \frac{\theta}{2}$.

To find the eigenvalues of $P_\lambda^2$ one needs a rotation to find
a harmonic oscillator and a free operator between $P_0,\;P_1$ and $P_2$.
The rotation is given by:
\bea\label{130}
P_1'=P_1\cos\phi  + P_2\sin\phi,\;\;\;P_2'=-P_1\sin\phi + P_2\cos\phi ,
\eea
with the condition $[P_0,P_2']=0$. In this way one finds that
$\tan\phi=\frac{\beta}{\alpha}$ and
\bea\label{135}
[P_0,P_1']=\frac{i}{2}\gamma\Sigma_3
\Rightarrow F_{01}'=\frac{-\gamma}{2}\Sigma_3,
\eea
in which $\gamma\equiv\sq{\alpha^2+\beta^2}$.
By making use of this rotation $F_{\mu\nu}$'s entries are then given by:
\bea\label{125}
a_1=\frac{1}{2}\gamma\Sigma_3,\;\;a_2=\cdots=a_5=0.
\eea
After some calculations one then obtains
\bea\label{136}
P'_2&=&\Delta P {\bf 1}_4,\;\;\;\;\;\;P*=[p,*], \nonumber\\
\Delta&=& \frac{\sqrt{\dpn}}{T}\frac{\sin\theta}{\gamma}
\left(-\frac{T^2L^2}{\dpn^2}+c^2\right)=
\frac{\sqrt{\dpn}}{T}\frac{\sin\theta}{\gamma}(-6\pi^2\ap^2),
\eea
and
\bea\label{137}
\gamma=2\sqrt{6}\pi\ap\sqrt{\cf^2+\sin^2\frac{\theta}{2}},\;\;\;\;
\Delta=-\frac{L}{\sqrt{\dpn}}\frac{\sin\theta}{2\sqrt{1+\cf^2}
\sqrt{\cf^2+\sin^2\frac{\theta}{2}}}.
\eea
The eigenvalues of $P_\lambda^2$ are found to be
\bea\label{140}
2\gamma(k+\frac{1}{2})+\Delta^2 \eta^2+r^2,
\eea
with $k$ as the harmonic oscillator number and $\eta^2$ as the
eigenvalues of the operator $P^2$. The one-loop effective action,
with a similar procedure of angled D-strings of \cite{IKKT}
reads therefore,
\bea\label{145}
W&\sim& -8N \left(\frac{\gamma}{2}\right)^3 \int \frac{d\eta}{\sqrt{\dpn}}
\frac{1}{(\Delta^2\eta^2+r^2)^3}=-8\sqrt\frac{N}{2\pi}\frac{3\pi}{16}
\left(\frac{\gamma}{2}\right)^3\frac{1}{|\Delta |r^5}
\nonumber\\
&\sim& -\frac{TL}{2\pi}3\pi(6\pi^2\ap^2)\frac{\left(\cf^2+
\sin^2\frac{\theta}{2}\right)^2}{\sin\theta}\frac{1}{Lr^5}.
\eea
This result is in agreement with the string theory one in the small
electric
field limit ($E,E'\ll 1$ in \cite{kaman})
\footnote{Since IIB matrix model is defined in Euclidean signature,
via a Wick rotation, one should replace $i\cf$ with $E$ and $E'$
in \cite{kaman}.}.


\noindent{\large\bf Acknowledgement}
\medskip

\noindent
Both authors thank  M.M. Sheikh-Jabbari for useful discussions, and N.
Sadooghi for reading the manuscript.

\end{document}